\pdfoutput=1
\documentclass{aip-cp}

\usepackage[numbers]{natbib}
\usepackage{rotating}
\usepackage{graphicx}

\newcommand{\ba}{\begin{eqnarray}}
\newcommand{\ea}{\end{eqnarray}}
\newcommand{\be}{\begin{equation}}
\newcommand{\ee}{\end{equation}}
\newcommand{\bmath}{\begin{mathletters}}
\newcommand{\emath}{\end{mathletters}}
\newcommand{\ban}{\begin{eqnarray*}}
\newcommand{\ean}{\end{eqnarray*}}

\newcommand{\bsub}{\begin{subequations}}
\newcommand{\esub}{\end{subequations}}

\begin{document}

\title{Van Hove Singularities and Excited-State Quantum Phase Transitions in Graphene-like Microwave Billiards}

\author[aff1]{Michal Macek\corref{cor1}}
\author[aff2]{Barbara Dietz}

\affil[aff1]{The Czech Academy of Sciences, Institute of Scientific Instruments, Brno, Czech Republic.}
\affil[aff2]{School of Physical Science and Technology, and Key Laboratory for Magnetism and Magnetic Materials of MOE, Lanzhou University, Lanzhou 730000, China}
\corresp[cor1]{Corresponding author: michal.macek@isibrno.cz}

\maketitle

\begin{abstract}
We discuss solutions of an algebraic model of the hexagonal lattice vibrations, which point out interesting localization properties of the eigenstates at van Hove singularities (vHs), whose energies correspond to Excited-State Quantum Phase Transitions (ESQPT).
We show that these states form stripes oriented parallel to the zig-zag direction of the lattice, similar to the well-known edge states found at the Dirac point, however the vHs-stripes appear in the bulk. We interpret the states as lines of cell-tilting vibrations, and inspect their stability in the large lattice-size limit.   
The model can be experimentally realized by superconducting 2D microwave resonators containing triangular lattices of metallic cylinders, which simulate finite-sized graphene flakes. Thus we can assume that the effects discussed here could be experimentally observed. 
\end{abstract}

\bigskip
{\it INTRODUCTION:}  
Excited state quantum phase transitions (ESQPTs)~\cite{ref:Cejn06,ref:Capr07} are a recent generalization of the quantum phase transitions (QPTs)~\cite{ref:Carr} and correspond to non-analytic behavior in excited spectra of low-dimensional systems. ESQPTs are currently studied in diverse systems including atomic nuclei, molecules, coupled atom-field quantum optics systems, driven quantum oscillators, and also two dimensional lattices, with seminal inputs in most of these fields by Franco Iachello. The interest in ESQPTs stems on one hand from the marked structural changes occurring in the individual systems, on the other hand from possible profound general implications for non-equilibrium thermodynamics, quantum information processing, and transport (for a review see~\cite{ref:CejnHERE}). 
A general classification of the ESQPTs is based on the dimensionality (number of degrees of freedom) of the system and the types of stationary points of the underlying semi-classical energy manifold, see Refs.~\cite{ref:esqptI,ref:esqptPLA}. Relation to stationary points may apparently lead to interesting localization effects at the ESQPT energies~\cite{ref:Hein06,ref:Lea15,ref:Crystals}.
\begin{figure}[h]
  \centerline{\includegraphics[width=\linewidth]{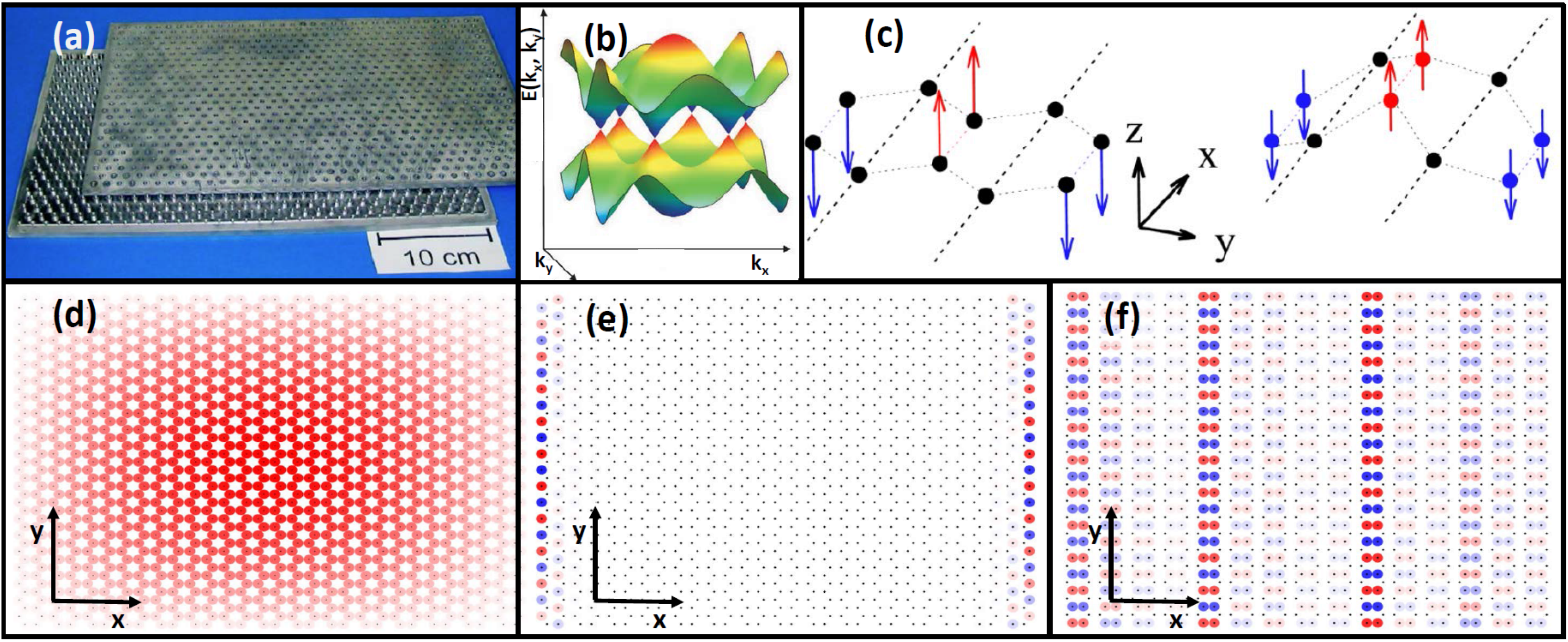}}
  \caption{Panel (a): Microwave resonator with rectangular boundaries containing approximately $900$ cylinders used in Darmstadt experiments, adapted from~\cite{ref:Diet15}. Panel (b): Energy dispersion relation $E(k_x,k_y)$, Eq.~(2), with $\lambda^{(I)}\neq 0$, $\lambda^{(II)}= 0$.  Panel (c): Schematic picture of the vHs- states formed by linear sequences of cell-tilting oscillations along the zig-zag direction with homogeneous oscillation amplitudes. Selected eigenstates of $30 \times 41$ hexagonal lattice with NN-hopping interactions ($\lambda^{(I)}\neq 0$, $\lambda^{(II)}= 0$): (d) ground state, (e) an edge state at the Dirac point and (f) a striped state at the lower vHs [cf. panel (c)]. Size of the points is proportional to the probability density of the $\nu$-th eigenstate $|\alpha_i^{(\nu)}|^2$ at site $i$, while the (red/blue) colors distinguish the (up/down) phases.}
\end{figure}
In this contribution, we concentrate on the localization of eigenstates in two dimensional lattices, motivated by a recent experimental identification of ESQPTs in artificial graphene simulated by microwave photonic crystals~\cite{ref:Diet13}. The experiments performed in Darmstadt~\cite{ref:Diet15} used microwave resonators with a set of cylinders in a triangular arrangement, representing nodes of the EM field, see Fig. 1 (a). The ESQPTs were observed in the density of states (DoS) at the energies of van Hove singularities (vHs) in one-phonon bands of the hexagonal lattice. Here, we present numerical solutions of an algebraic model of the system~\cite{ref:Iach15}. Complementing Ref.~\cite{ref:Crystals}, which showed that eigenstates at the ESQPT/vHs energy localize into peculiar stripes oriented parallel to the zig-zag direction of the lattice, we interpret these states structurally as linear sequences of cell-tilting vibrations. Further, we inspect their stability in the large lattice-size limit.

{\it MODEL:}
We consider the ``hopping'' limit of the algebraic Hamiltonian introduced by Franco Iachello~\cite{ref:Iach15}
\ba
\label{eq:H}
H=\varepsilon \sum_{i=1}^{n} b_{i}^{\dag }b_{i}&-&\lambda
^{(I)}\sum_{\langle i,j\rangle}\left( b_{i}^{\dag }b_{j}+b_{j}^{\dag }b_{i}\right) - \lambda ^{(II)}\sum_{\langle\!\langle i,j\rangle\!\rangle}\left( b_{i}^{\dag }b_{j}+b_{j}^{\dag
}b_{i}\right) +... 
\ea
where the hopping (Majorana) operator on the hexagonal lattice is expanded into the nearest neighbors (interaction strength coefficient $\lambda ^{(I)}$), the next-to-nearest neighbors (coefficient $\lambda ^{(II)}$), etc..., terms. The coefficients $\lambda ^{(I)},\lambda ^{(II)},...$ are symmetry adapted to reflect the hexagonal unit cell. 

For an infinite size hexagonal lattice, the energy dispersion relation (EDR) can be written as~\cite{ref:Iach15}
\begin{eqnarray}\label{eq:HexagonEDRcont}
E\left( k_{x},k_{y}\right) &=&\pm \lambda ^{(I)}\sqrt{3+u\left(
k_{x},k_{y}\right) }-\lambda ^{(II)}u\left( k_{x},k_{y}\right)  \nonumber \\
u\left( k_{x},k_{y}\right) &=&2\cos 2\pi k_{y}a+4\cos \pi k_{y}a\cos \pi
k_{x}\sqrt{3}a
\end{eqnarray}%
where $\pi k_{x} a$ and $\pi k_{y} a$ are real quasimomenta, ranging from $0$ to $\pi$. The two sheets of the energy surface, Eq.~(\ref{eq:HexagonEDRcont}), of the infinite-size hexagonal lattice touch each other conically at Dirac points at the corners of the first Brillouin zone, see Fig. 1(b). The density of states (DoS) corresponding to Eq.~(\ref{eq:HexagonEDRcont}) can be written in terms of elliptic integrals~\cite{ref:Iach15}. 
The DoS shows two van Hove singularities (vHs), related to the saddle points of the EDR, Eq.~(2). In the infinite lattice size limit, the DoS diverges logarithmically at the vHss and vanishes at the Dirac point (or Dirac ``zero'', Dz).

\begin{figure}[h]
  \centerline{\includegraphics[width=\linewidth]{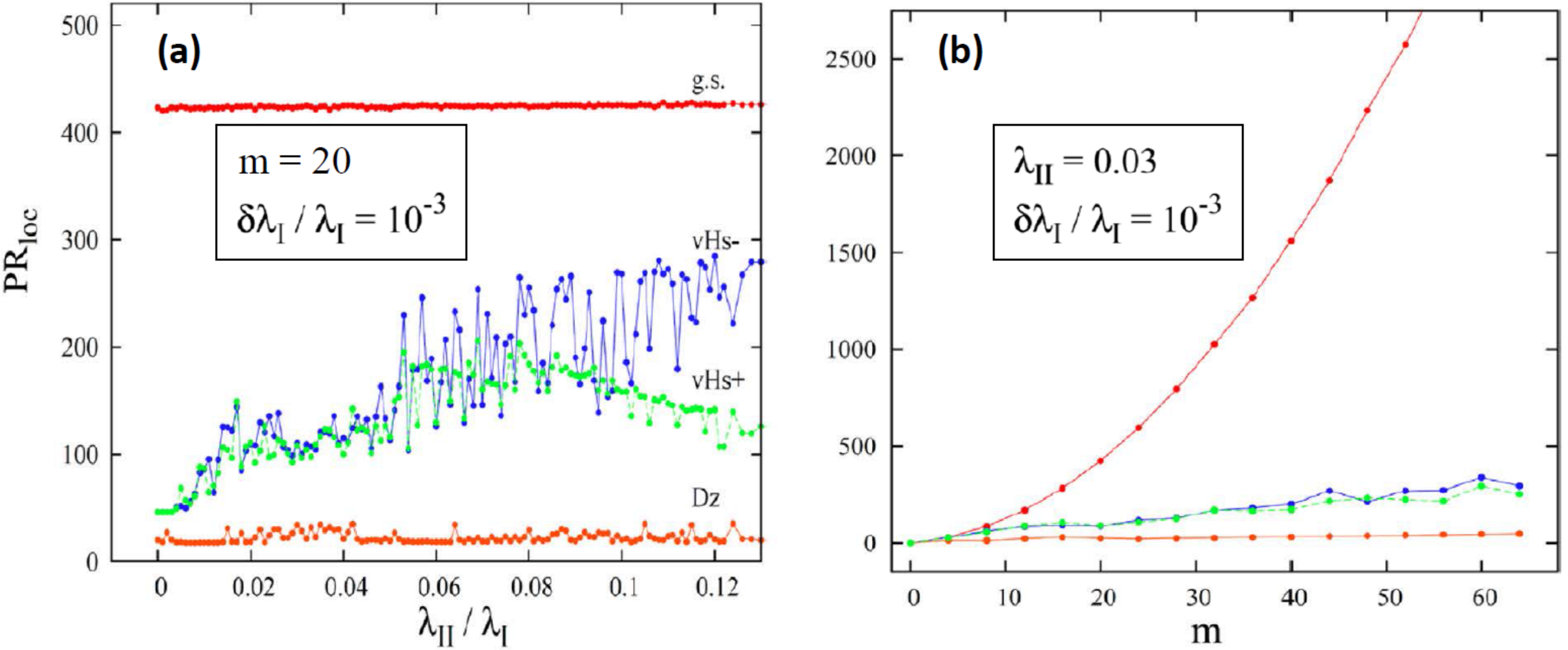}}
  \caption{Evolution of the local-basis participation ratio PR$_\mathrm{loc}^\nu \equiv \sum_i 1/|\alpha_i^{(\nu)}|^4$ for the ground state (g.s.) and the states at the Dirac zero (Dz) and both vHss (vHx$-$, vHs$+$) as a function of the relative strength $\lambda_{II}/\lambda_{I}$ of first and second-neighbor interaction (panel a) and the size of the lattice with $m\times (2m+5)$ sites (panel b). The coefficients are affected by noise uniformly distributed in respective intervals $|\delta\lambda_i/\lambda| \leq 10^{-3}$ in both panels.}
\end{figure}

{\it LOCALIZATION OF WAVE FUNCTIONS:}
Selected eigenstates of the hexagonal lattice with $\lambda^{(I)}\neq0$, and $\lambda^{(II)}=0$ in different regions of the one-phonon band are presented in Fig. 1 (panels d-f). These were obtained by numerical diagonalization of the Hamiltonian (\ref{eq:H}) on a finite lattice with $30 \times 41$ sites with open (Dirichlet) boundary conditions (the $x$-axis is parallel to the armchair direction, while the $y$-axis to the zig-zag direction). The probability density $|\alpha_i^{(\nu)}|^2$ of the $\nu$-th eigenstate at the site $i$ is visualized by circles of radius proportional to $|\alpha_i^{(\nu)}|^2$, while red/blue color indicate the positive/negative phase. The ground state (panel d), as well as all the states (not shown) below the lower vHs display a simple sinusoidal waveform resembling vibrations of a continuous membrane. At the Dirac point (panel e), we observe the markedly different edge states along the zig-zag edge, which are of topological origin and relate to the two separate triangular sub-lattices of the hexagonal lattice~\cite{ref:Hats02}. Most interestingly, at both van Hove singularities, the eigenstates form stripes (panel f shows a state at the lower vHs, cf.~\cite{ref:Crystals}). Similar to the edge states, they are oriented parallel to the zig-zag direction, the vHs-stripes however appear in the bulk. Inspecting the phases and amplitudes of the striped states, we can see that the vHs states correspond to specific linear sequences of unit-cell-tilting vibrations. This is schematically depicted in panel (c). At the upper van Hove singularity vHs$+$, the states form similar stripes which differ only by phases. As a consequence, a unit-cell-twisting vibration occurs (not shown). We point out that the eigenstates at both vHss have nodal lines parallel to the armchair direction and the number of striped states observed at each vHs equals half the number of zig-zag rows in the lattice. 

Second-neighbor interaction $\lambda^{II}\neq 0$ has a strong delocalizing effect on the striped eigenstates at the vHs. However the localization can be recovered if weak noise is introduced, meaning that $\lambda^{I}$ and/or $\lambda^{II}$ are not constant throughout the lattice, cf.~\cite{ref:Crystals}. Fig. 2(a) shows the evolution of the local-basis participation ratio PR$_\mathrm{loc}^\nu \equiv \sum_i 1/|\alpha_i^{(\nu)}|^4$ for the ground state (g.s.) and the most localized eigenstates at Dz and both vHss (vHs$+$ and vHs$-$), as a function of $\lambda^{II}/\lambda^{I}$ for a hexagonal lattice with $m \times (2m+5)$ sites, where $m=20$. The coefficients $\lambda^{I}$ and $\lambda^{II}$ are affected by noise $\delta\lambda^{I,II}_i$ uniformly distributed in respective intervals $|\delta\lambda^{I,II}_i/\lambda^{I,II}| \leq 10^{-3}$.
The participation ratio PR$_\mathrm{loc}^\nu$ counts, roughly speaking, the sites where $|\alpha_i^{(\nu)}| > 0$, thus the lower the PR$_\mathrm{loc}^\nu$, the more localized the $\nu$-th eigenstate. We see that at $\lambda_{II}/\lambda_{I}\approx 0$, the PR$_\mathrm{loc}$-values for both the vHs states are roughly $2\times$ larger than for the Dirac-point state, while being about $10\times$ lower than for the ground state. This is consistent with spatial distributions shown in Fig. 1(d-f). As $\lambda_{II}/\lambda_{I}$ increases, both vHs states gradually delocalize up to $\lambda_{II}/\lambda_{I}\approx 0.06$, while the edge states at Dz remain essentially unaffected. Dependencies of PR$_\mathrm{loc}$ on the lattice size $m\times (2m+5)$ shown in Fig. 2 (b) prove that eigenstates at both vHss and Dz are localized on linear portions of the lattice as PR$_\mathrm{loc} \propto m^1$, in sharp contrast with the g.s. spread over the whole lattice expressed by the quadratic scaling PR$_\mathrm{loc} \propto m^2$. Thus we have shown that the localization of the eigenstates of the hexagonal lattice into striped states can be robust and we suggest that such effects could be experimentally observed in an apparatus similar as in the detection of the edge states~\cite{ref:Diet15}.

The authors would like to thank Franco Iachello for all his support during the years, and for initiating and collaborating on this project. Work with him has been an honor, immense chance to learn and great pleasure! 


\nocite{*}
\bibliographystyle{aipnum-cp}%

\end{document}